\documentclass[11pt,twoside]{article}

\usepackage{asp2006}
\usepackage{epsf}
\usepackage{psfig}
\usepackage{lscape}

\begin{document}
\title{Magnetic Fields and Star Formation in Galaxies of different morphological types}
\author{Marita Krause}
\affil{Max-Planck-Institut f\"ur Radioastronomie, Auf dem H\"ugel
69, 53121 Bonn, Germany}

\begin{abstract}
From our radio continuum and polarization observations of a sample of
spiral galaxies with different morphological types, inclinations, and
star formation rates (SFR) we found that galaxies with low SFR have
higher thermal fractions/ smaller synchrotron fractions than those
with normal or high SFR.  Adopting an equipartition model, we
concluded from our observations that the nonthermal radio emission and
the \emph{total magnetic field} strength grow nonlinearly with SFR.

We also studied the magnetic field structure and disk thicknesses in
highly inclined (edge-on) galaxies. We found in five galaxies that
-despite their different radio appearance- the vertical scale heights
for both, the thin and thick disk/halo, are about equal (0.3/1.8kpc),
independently of their different SFR. They also show a similar
large-scale magnetic field configuration, parallel to the midplane and
X-shaped further away from the disk plane, independent of Hubble type
and SFR in the disk. Hence we conclude that the amplification and
formation of the \emph{large-scale} magnetic field structure is
independent of SFR.
\end{abstract}

\section{Observations of magnetic fields}

Radio observations of the continuum emission in the cm-wavelength
regime are the best way to study magnetic fields in galaxies.
Magnetic fields consist of regular and turbulent components. The total
magnetic field strength in a galaxy can be estimated from the
nonthermal radio emission under the assumption of equipartition
between the energies of the magnetic field and the relativistic
particles (the so-called {\em energy equipartition}) as described in
Beck \& Krause (2005).\\
The linear polarization reveals the strength and structure of the resolved regular magnetic field. Its analysis and the observational results of the field strengths and patterns of nearby face-on galaxies are summarized e.g. by Beck (this volume).\\

\section{Total magnetic field strength and star formation}

Observations of a sample of three late-type galaxies with low surface-brightness and the radio-weak edge-on galaxy NGC~5907 (all with a low SFR) revealed that they all have an unusually high thermal fraction and weak total and regular magnetic fields (Chy{\.z}y 2007, Dumke et al. 2000). However, these objects still follow the total radio-FIR correlation, extending it to the lowest values measured so far. Hence, these galaxies have a lower fraction of synchrotron emission than galaxies with higher SFR. It is already known that the thermal intensity is proportional to the SFR. Our findings fits to the equipartition model for the radio-FIR correlation (Niklas \& Beck 1997), according to which the nonthermal emission increases $\propto SFR^{\approx 1.4}$ and the \emph{total} magnetic field strengths as well as the ratio of nonthermal-to-thermal emission increases with $\propto SFR^{\approx 0.4}$.

\section{Regular magnetic field strength and star formation}

Several edge-on galaxies of different Hubble type and covering a wide range in SFR were observed with high sensitivity in radio continuum and linear polarization.
These observations show that the magnetic field structure is mainly {\em parallel to the disk} along the midplane of the disk (with the only exception of NGC4631) as expected from observations of face-on galaxies and their magnetic field amplification by the action of a mean-field $\alpha\Omega$-dynamo. Away from the disk the magnetic field has also vertical components increasing with distance from the disk and with radius (Krause 2004, Soida 2005, Krause et al. 2006, Heesen et al. submitted). Hence, the large-scale magnetic field looks X-shaped away from the plane and can be explained by a dynamo action together with a galactic wind (Brandenburg et al. 1993, Heesen et al. submitted).

We also determined the exponential scale heights for those five edge-on galaxies (NGC~253, NGC~891, NGC~3628, NGC~4565, NGC~5907) for which we have combined interferometer and single-dish data (VLA and the 100-m Effelsberg) at 6cm. In spite of their different intensities and extents of the radio emission, the scale heights of the thin disk and the thick disk/halo are similar in this sample (300pc and 1.8kpc) (Dumke \& Krause 1998, Dumke et al. 2000, Heesen et al.  submitted). We stress that our sample includes the brightest halo observed so far, NGC~253 with a very high SFR as well as one of the weakest halos, NGC~5907 with a small SFR. Hence we conclude that -though a high SFR in the disk increases the total magnetic field strength in the disk and halo-  it cannot change the global field configuration nor influences the global scale heights of the radio emission. The similar scale heights and magnetic field configurations imply similar propagation velocities of the cosmic ray electrons in these galaxies.

\end{document}